\documentclass[a4paper]{article}

\usepackage[english]{babel}
\usepackage[utf8]{inputenc}
\usepackage{fullpage}
\usepackage{amsmath}
\usepackage{graphicx}
\usepackage{pst-all}
\usepackage[colorinlistoftodos]{todonotes}
\usepackage{setspace}
\usepackage{pstricks}
\usepackage{bbold}
\usepackage{caption}
\usepackage{subcaption}
\usepackage{pict2e}

\providecommand{\e}[1]{\ensuremath{\times 10^{#1}}} 

\title{The Quantum Wavefunction}

\author{Nagaganesh Jaladanki}

\begin{document}
\maketitle

\onehalfspacing

\begin{abstract}
When most people think of physics, they think of what they learned in high school physics: that the world is fundamentally predictable.  Given the position and velocity of a particle in space, it should be possible to predict its position at any moment in the future---right? Though this was thought to be true for thousands of years, recent developments in the field of physics have shown that this isn't actually true. Instead of being fundamentally predictable, the universe is fundamentally  \textit{unpredicable}. However, this doesn't seem to make sense. What happened to the centuries of physics developed by Newton, Bernoulli, and Lagrange? Well, as it turns out, they weren't actually wrong. Their equations were actually an approximation of a formula called the wavefunction, which is the ``Newton's Laws'' equivalent for modern physics. In this paper, we'll take a look at this peculiar wavefunction---and why our intuition about the world is completely wrong.

\centerline{This is Quantum Physics.}

\end{abstract}

\section{Introduction and Background}

\subsection{Quantum vs. Classical}

Let's think back to our example of a particle moving through space. This particle would have (among many) two properties that we would be able to measure: its position $\vec{x}$ in space, and its momentum $\vec{p}$.

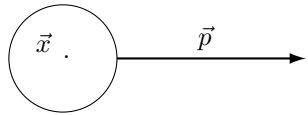
\begin{figure}[h]
\setlength{\unitlength}{0.14in}
\centering
\begin{picture}(32,6)
\put(13,3){\circle{4}}
\put(13,3){.}
\thicklines
\put(15,3){\vector(1,0){7}}
\put(12,3.25) {$\vec{x}$}
\put(18,3.5){$\vec{p}$}
\end{picture}

\caption{Because we need a figure for everything.}
\label{fig:ball}
\end{figure}

In classical physics, it would have been possible to know both of these values with exact precision. And we could then have coupled this information with the (classical) laws of physics to predict the future of this particle at any moment. If we think about this for a moment, it seems to make sense. If we release a box in deep space at some velocity, we would expect it to stay at that velocity for millions of years, as there is nothing that would block its movement. As a result, we can use some simple equations to predict its exact movement and position.

This, essentially, forms the backbone of classical physics. If we know the state of the universe at a single moment in time, then we can use the laws of physics to predict the future. This is known as determinism---the idea that everything can be determined from an initial value. And this is exactly what quantum physics breaks.

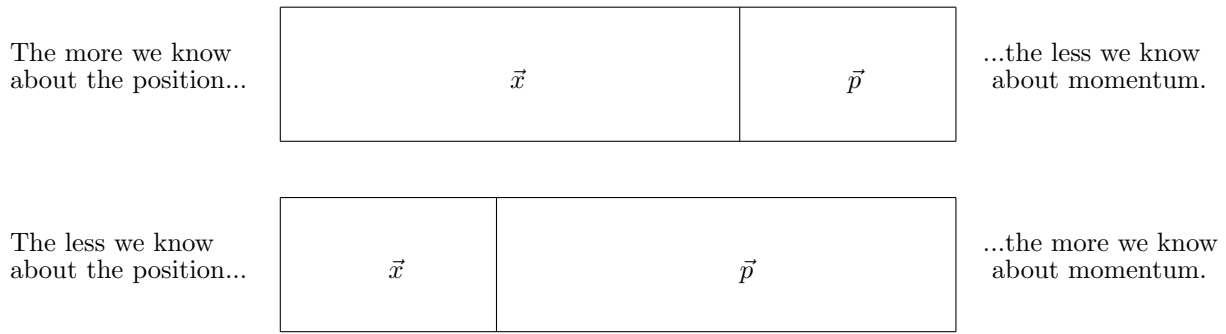
\begin{figure}[h]
\setlength{\unitlength}{0.14in}
\centering

\begin{picture}(45,8)
\put(0,5){The more we know}
\put(0,4){about the position...}
\put(36,5){...the less we know}
\put(36,4){ about momentum.}
\put(10,2){\line(0,1){5}}
\put(10,7){\line(1,0){25}}
\put(35,7){\line(0,-1){5}}
\put(35,2){\line(-1,0){25}} 
\put (27,2){\line(0,1){5}}
\put(18.5,4) {$\vec{x}$}
\put(31,4){$\vec{p}$}
\end{picture}

\begin{picture}(45,5)
\put(0,3){The less we know}
\put(0,2){about the position...}
\put(36,3){...the more we know}
\put(36,2){ about momentum.}
\put(10,0){\line(0,1){5}}
\put(10,5){\line(1,0){25}}
\put(35,5){\line(0,-1){5}}
\put(35,0){\line(-1,0){25}} 
\put (18,0){\line(0,1){5}} 
\put(14,2) {$\vec{x}$}
\put(27,2){$\vec{p}$}
\end{picture}

\caption{The Uncertainty Principle}
\label{fig:uncertain}
\end{figure}

Here are the bare facts: quantum physics states that there is a limit to how precisely we can measure the position and momentum of anything in the universe. This is known as the Heisenberg Uncertainty Principle. It says that the error of measurement of the position and momentum have to be greater than a specific (though \textit{very} small) value. In this case, the term ``error of measurement'' doesn't  concern our ability to measure these properties---the Uncertainty Principle is true no matter what device we use to measure position and momentum. As a result, the more specifically we know position, the less sure we can be for the particle's momentum. Similarly, the more information we have about the particle's momentum, the more uncertain we are about its position. You can see this in Figure~\ref{fig:uncertain}.

To understand the Uncertainty Principle, suppose that in a special deal, a restaurant is offering an all-you-can-eat for one plate of food. There are two separate piles of food, which correlate to the lists A and B. At the restaurant, all of the food that a customer takes can be from pile A, or it can be from pile B. This is comparable to knowing exactly where a particle is, or knowing exactly its momentum. However, the customer can also choose to fill half of his plate with items from pile A, and half with items from pile B. In this case, the scientist would know with only some certainty on where the particle is and some certainty on the particle's momentum.

One could ask: why don't we see this in everyday life? It doesn't take much effort to find the position \textit{and} the momentum of a car on the highway. The answer is that the objects that we handle every day are so big and full of energy (compared to the quantum world) that the effects of quantum physics are barely measurable. Where are they measurable? Let's see in a small example.

Imagine the same scenario as Figure~\ref{fig:ball}, but replace the ball with an electron. And, being the scientists that we are, we decide to measure the electron's position and momentum. Because of the Uncertainty Principle, we can't be completely certain of both position and momentum. This is how we can visualize it:

\begin{figure}[h]
\includegraphics[width=0.4\textwidth]{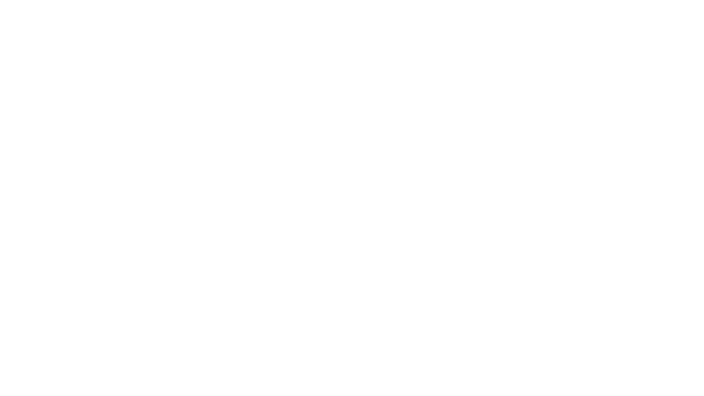}
\caption{Electrons and Uncertainty}
\label{fig:electron}
\end{figure}

We can't. This is because the electron is very, \textit{very}, small, and the levels of uncertainty are much bigger than the size of the electron. ``Small'' is a word that quantum physics likes, as most of its effects are apparent at these levels. Consequently, we can visualize a car or a box very easily, but not an electron. However, this doesn't mean that the electron doesn't exist. It still exists, but may be anywhere within the uncertainty bound---which will be very high for an electron.

That's a curious thing to think about: the fact that the electron, a particle, can be anywhere in a wide region of space or have a range of values for momentum. At this point, the electron doesn't seem to act like a particle anymore.

\subsection{The Wave-Particle Duality}
\label{waveparticle}

When we said that the particle could be anywhere within the bound of uncertainty when we measured it, we were actually lying. The truth isn't that the particle could be \textit{anywhere} within the bound. It is actually \textit{everywhere}  within the bound. Yet at the same time, it's still a particle. This idea is what led to the creation of the Wave-Particle Duality. 

The Wave-Particle Duality states that every particle acts as both a particle and a wave.  To understand this a little better, think of a particle as a cylinder. There are two ways of looking at it. Viewed from the top, it appears as a circle, while a side view paints a rectangular picture. Likewise, we can think of a particle as having the properties of both a particle (intuitive) and a wave (unintuitive).

So far, you've been taking the things that I've been saying as true. But we need to be grounded in reality, so let's take a look at an experiment that provides a real life basis for all this theory we've been talking about.

\begin{figure}[h]
\setlength{\unitlength}{0.14in}
\centering
\begin{picture}(40,10)
\put(35,10){\line(0,-1){10}}
\put(15,10){\line(0,-1){3.5}}
\put(15,5.75){\line(0,-1){1.5}}
\put(15,3.5){\line(0,-1){3.5}}
\put(2,5.75){\line(0,-1){1.5}}
\put(2,4.25){\line(1,0){3}}
\put(5,4.25){\line(0,1){1.5}}
\put(2,5.75){\line(1,0){3}}
\thicklines
\put(5,5){\vector(1,0){4}}
\put(5,5){\vector(4,1){4}}
\put(5,5){\vector(4,-1){4}}
\end{picture}
\caption{The Experimental Setup.}
\label{fig:youngapparatus}
\end{figure}
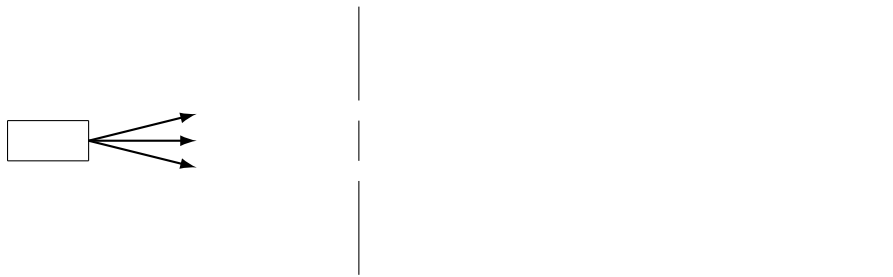

The experiment is as follows: a stream of very small particles (such as electrons) are fired at a small barrier with two narrow slits in it. The particles that go through the slit are then recorded on a particle detector that lies far from the detector. The experimental apparatus is shown in Figure~\ref{fig:youngapparatus}.\footnote{This experiment is commonly known as Young's experiment.}

With this figure, it's easy enough to see that the detector will have two broad stripes corresponding for each of the slits. This is true for most objects that we can see. For example, consider that we were hurling baseballs to go between the slits. Most of the baseballs will bounce back after missing the slit. However, the few that make it through will hit the detector in one of the two big stripes.

However, once again, this doesn't hold true for very small particles, such as electrons or photons. Instead of two stripes, we see a very interesting pattern, which you can see in Figure~\ref{fig:young_prob}.

\begin{figure}[h]
\centering
\includegraphics[width=0.6\textwidth]{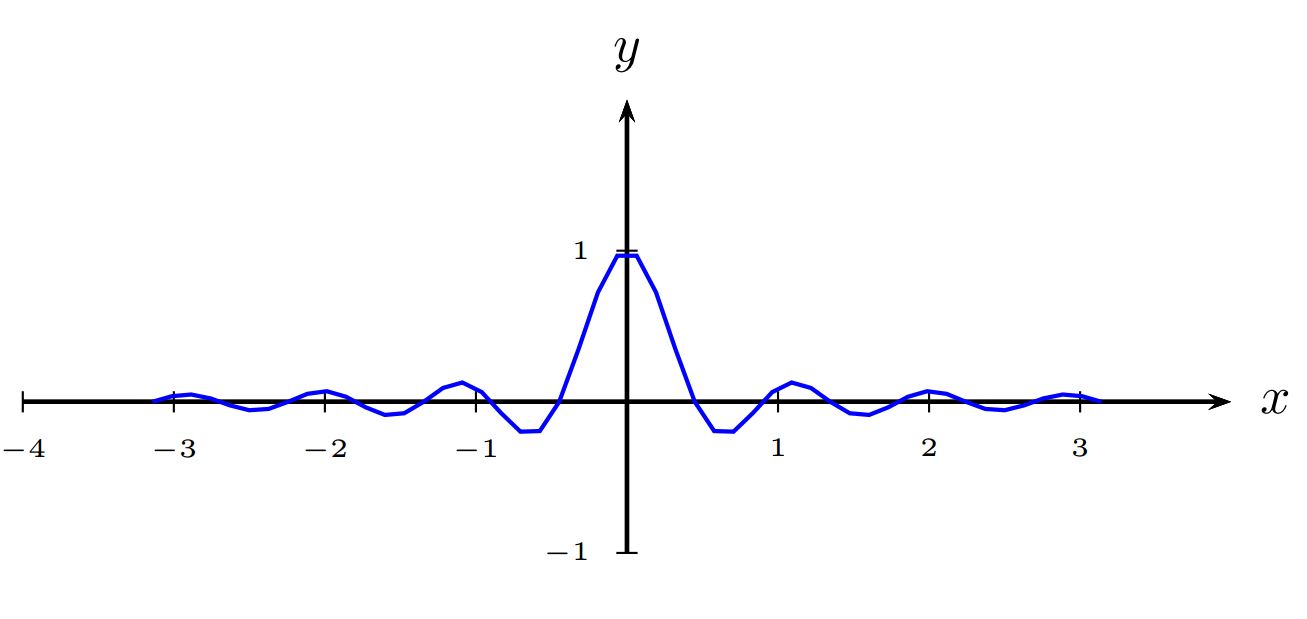}
\caption{The Experiment for Waves}
\label{fig:young_prob}
\end{figure}

In Figure~\ref{fig:young_prob}, the $y$-axis refers to the point on the detector which is exactly between the two slits, and the $x$-axis lies along the detector.\footnote{This graph is just an approximation to the actual, more complicated data. However, the essential features have remained unchanged.} In addition, the crests represents areas where many particles hit the detector, and the troughs show the areas which were hit relatively fewer times. Even at first glance, we can tell that something is pretty wrong here. For example, why are the particles hitting the detector at multiple spots? And how do they hit the detector so far from the origin? Both of these questions are answered by the Wave-Particle Duality. 

\begin{figure}[h]
\centering
\includegraphics[width=0.5\textwidth]{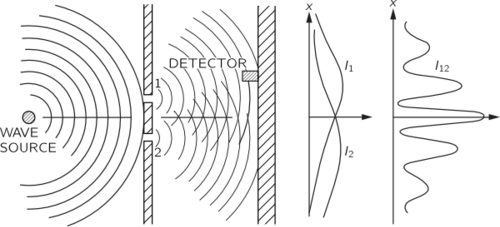}
\caption{Particles in the Experiment.}
\label{fig:young_experi}
\end{figure}

As they say, a picture is worth a thousand words. Take a look at Figure~\ref{fig:young_experi} for a visual. As you can see, the particles are behaving as waves when they pass the slit. The waves then interfere, producing constructive and destructive interference. By the time the particles reach the detector, they have interfered greatly, producing the unintuitive pattern that we see in Figure 5. However, that's not all there is to it. It's important to note that different electrons are \textit{not} interfering with each other---each electron is interfering with itself.

Each electron is producing a wave that passes through both slits, interferes with itself, and lands on the detector in one spot. As a result, this pattern would appear even if each particle was released separately. So far, we've been dealing with the theory of quantum physics. Let's make this a little more concrete. But how do we do that?

Math is the answer. (It always is.) We can quantitatively describe what is happening to the particles when they become waves by graphing their wave function. How do we get ahold of this wavefunction?

\begin{centering}
Behold the greatest formula of our time. Schr{\"o}dinger's equation.

\[ \scalebox{1.5}{$E\Psi(x)=U\Psi(x)-\frac{\hbar}{2m} \frac{d^2\Psi(x)}{dx^2}$} \] 

\end{centering}

\section{The Holy Wavefunction}
\subsection{What is the Wavefunction?}

Though they seem similar, the wavefunction and Schr{\"o}dinger's equation are separate things. The wavefunction describes the Wave-Particle action of a system (or a particle), while the Schr{\"o}dinger equation gives you a way to generate the wavefunction of a particular system.

Before we take a look at why Schr{\"o}dinger's equation makes sense, let's take a look at what it's trying to say. 

\[ \scalebox{1.2}{$E\Psi(x)=U\Psi(x)-\frac{\hbar}{2m} \frac{d^2\Psi(x)}{dx^2}$} \] 

\begin{figure}[h]
\setlength{\unitlength}{0.14in}
\centering
{$E$ = Energy in the system}\\*
{$\Psi(x)$ = The Wave Function of the System}\\*
{$U$ = The potential energy}\\*
{$\hbar$ = The reduced Planck constant}\\*
{$m$ = The mass of the system}\\*
{$\frac{d^2\Psi(x)}{dx^2}$ = The second derivative of the wavefunction.}
\caption{Understanding Schr{\"o}dinger's equation.}
\label{fig:schrodinger}
\end{figure}

Most equations are like a map. Schr{\"o}dinger's equation is as well. You can use the legend in Figure~\ref{fig:schrodinger} to understand what the symbols in this equation mean. As you may have seen, many of the symbols in this equation are simple symbols from classical physics.

In this equation, the letter $\Psi$ (pronounced \textit{sigh} and written as \textit{Psi}) represents the wavefunction.

Now, let's take a look at how this equation is consistent with some previous examples of physics, such as the conservation of energy and other properties.

\subsection{The Phony Derivation}

Disclaimer: This isn't actually a derivation. Schr{\"o}dinger's equation is actually a fundamental principle of physics, so it isn't actually possible to ``derive'' it using other principles. Instead, we will merely show that it is compatible with the laws of physics.

First, let's establish some basic principles that we will modify later on. To begin with, the wavelength $\lambda$ of a particle (due to the Wave-Particle Duality), is:

\begin{equation} \label{eq:debroglie}
\lambda = \frac{h}{p}
\end{equation}

where $h$ is Planck's constant and $p$ the magnitude of the momentum.

We also know from the Wave-Particle duality that particles act like waves---something like a pendulum moving back and forth. How do we represent a wave mathematically? Using a differential equation:

\begin{equation} \label{eq:wave}
\frac{d^2\Psi(x)}{dx^2} = -k^2\Psi(x)
\end{equation}

For this wave, $\Psi$ represents the wavefunction (as we have seen that it acts like a simple wave) and $k$ represents the wavenumber, or ``how quickly the wave moves''. The variable $k$ is defined to be:

\begin{equation} \label{eq:wavenumber}
k = \frac{2\pi}{\lambda}
\end{equation}

Almost immediately, we can see that we can shift some variables around to get a slightly diferent equation. Combining Equation 3 with Equation 1, we can get:

\begin{equation}
k = \frac{p}{\hbar}
\end{equation}

where $\hbar$ is just $\frac{h}{2\pi}$. We seem to be getting somehwere. Replacing $k$ in Equation 2 and doing some rearranging,
\begin{subequations}
\begin{equation}
\frac{d^2\Psi(x)}{dx^2} = -\frac{p^2\Psi(x)}{\hbar^2}
\end{equation}

\begin{equation}
-\frac{d^2\Psi(x)}{dx^2} \hbar^2  = p^2\Psi(x)
\end{equation}
\end{subequations}

Our equation is taking form from the symbols. To get further, we decide to divide both sides by $2m$. 

\begin{equation}
-\frac{d^2\Psi(x)}{dx^2} \frac{\hbar^2}{2m}  = \frac{p^2}{2m}\Psi(x)
\end{equation}

Now, remember that:

\begin{equation}
\frac{p^2}{2m} = K
\end{equation}

where $K$ is the kinetic energy in the system.\footnote{Assuming nonrelativistic speeds.} This leads to:

\begin{equation} \label{eq:almostthere}
-\frac{d^2\Psi(x)}{dx^2} \frac{\hbar^2}{2m}  = K\Psi(x)
\end{equation}

Let's relate this to the Principle of Conservation of Energy, which states that the total energy in a system is equal to the kinetic energy added to the potential energy.

\begin{subequations}
\begin{equation}
E = K + U
\end{equation}

\begin{equation}
K = E - U
\end{equation}
\end{subequations}

Plugging this into Equation 8 and rearranging, we get:

\begin{subequations}
\begin{equation}
-\frac{d^2\Psi(x)}{dx^2} \frac{\hbar^2}{2m}  = (E-U)\Psi(x)
\end{equation}

\begin{equation}
-\frac{d^2\Psi(x)}{dx^2} \frac{\hbar^2}{2m} + U\Psi(x) = E\Psi(x)
\end{equation}

\begin{equation}
E\Psi(x) = U\Psi(x) - \frac{d^2\Psi(x)}{dx^2} \frac{\hbar^2}{2m}
\end{equation}
\end{subequations}

Voila! Our equation is identical to Schr{\"o}dinger's equation. Let's take a look back at what we've done. We've shown that Schr{\"o}dinger's equation is compatible with Conservation of Energy and the Wave-Particle Duality. Nevertheless, this is still a phony derivation. Schr{\"o}dinger's equation is a fundamental aspect of nature, so it isn't possible to derive it---you can only show that it's compatible with other fundamental principles of nature, which is what we've shown. The version that we've just shown is true is known as the one-dimensional time-independent Schr{\"o}dinger equation. 

We now know why Schr{\"o}dinger's equation is what it is. However, what does it mean?

\subsection{Remarks}
Let's take a few moments to understand what exactly $\Psi(x)$ is. We've already established that it represents the wave that particle acts as. However, we can ask: what does the wave displace? Water waves displace water, and sound waves displace pressure. Unfortunately, the waves that $\Psi(x)$ represents don't displace anything. In fact, they don't even exist. The wave function acts as a mathematical representation of what happens in reality. If we base this off of experimental data, we find that the particles merely \textit{act} as waves. No one knows whether they are or aren't.

Though the wave that the wavefunction represents doesn't exist, the square of the wavefunction has a physical meaning. The square of the wavefunction at a particular location represents the probability that a particle will be at that location.

This makes a little more sense in context. Here is an example of a probability density function derived from the wavefunction:

\begin{figure}[h]
\centering
\includegraphics[width=0.6\textwidth]{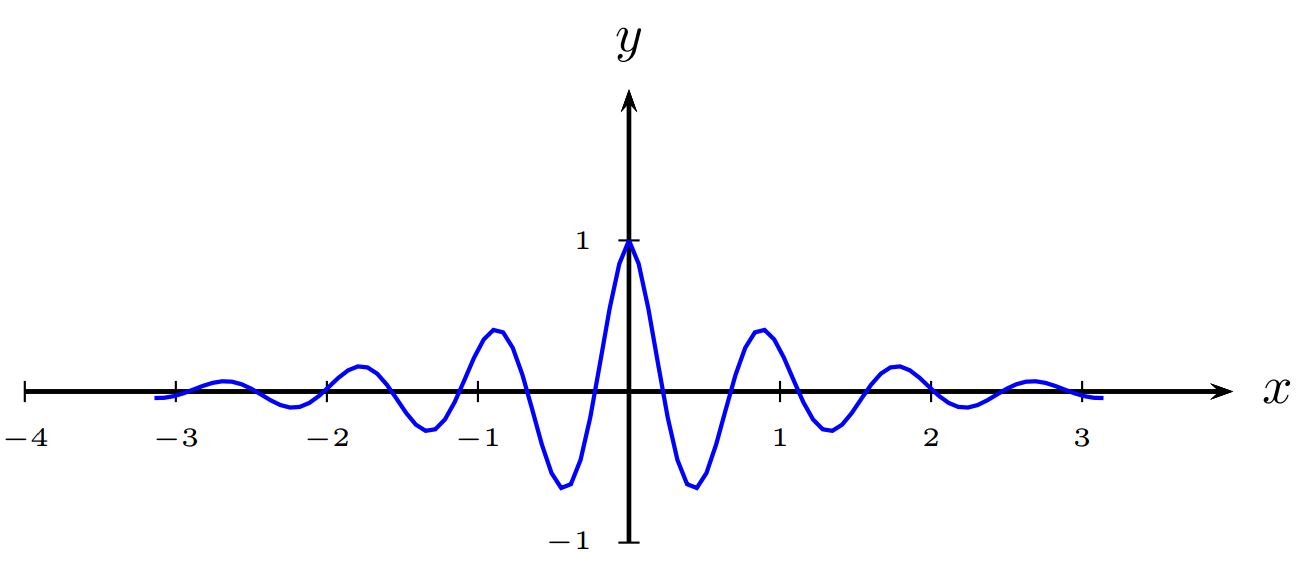}
\caption{The Experiment for Waves}
\label{fig:young_prob}
\end{figure}

It looks familiar, as it's merely the square of the wave function for the Wave-Particle Duality experiment. However, here's an example of how it looks when the experiment is actually done.

\begin{figure}[h]
\centering
\includegraphics[width=0.5\textwidth]{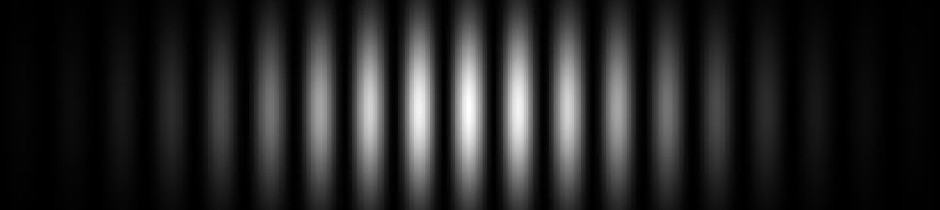}
\caption{Experimental Results.}
\label{fig:young_experi_2}
\end{figure}

We can see that the wavefunction matches up with the experimental results, which is good. However, let's trace the path of one particle as it goes through the experiment. As we've seen above, the particle acts as a wave until it reaches the barrier. The instant it reaches the barrier, the wave picks a position based on probability ($\Psi^2(x)$) and collapses into a particle, which the detector records. This collapse is the reason why we can't measure the actual wave in a laboratory experiment---we can only see the effects that it has.

\section{Schr{\"o}dinger's Equation Applied}\
In this section, we'll go over an example of Schr{\"o}dinger's equation in action. In addition, we'll also assume a little more math and physics knowledge to speed up the process. 

\subsection{Particle in an Infinitely Deep Well}
In this example, we'll be analyzing an electron trapped in a potential well, so that its potential energy function something looks like this:

\begin{figure}[h]
\centering
\includegraphics[width=0.6\textwidth]{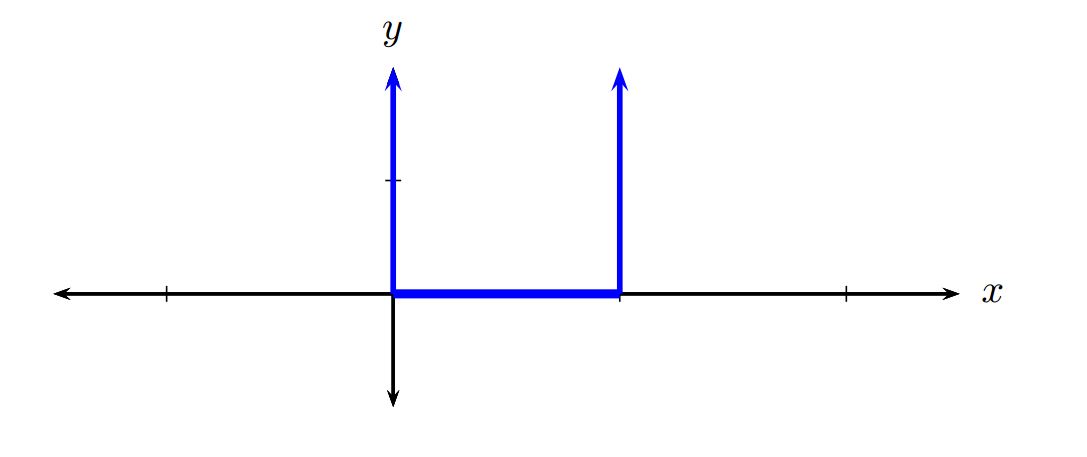}
\caption{Infinitely Deep Potential Well}
\label{fig:young_prob}
\end{figure}

We're also given that the electron is trapped between the points 0 and $L$. It's important to note that the electron exists only in one dimension---it's just the potential energy that differs between various points in space.

Now that we've got the description out of the way, let's try to to apply Schr{\"o}dinger's Equation to this situation.

\begin{equation}
E\Psi(x) = U\Psi(x) - \frac{d^2\Psi(x)}{dx^2} \frac{\hbar^2}{2m}
\end{equation}

We know that the potential energy at any point where the particle can exist must equal zero, as shown by the graph. As a result, 

\begin{subequations}
\begin{equation}
E\Psi(x) =  - \frac{d^2\Psi(x)}{dx^2} \frac{\hbar^2}{2m}
\end{equation}
\begin{equation}
-\frac{2Em}{\hbar^2}\Psi(x) = \frac{d^2\Psi(x)}{dx^2}
\end{equation}
\end{subequations}

This equation is really similar to Equation~\ref{eq:wave}, and in fact shares the same general solution.

\begin{equation}
\Psi(x) = A\sin(kx) + B\cos(kx)
\end{equation}
where $A$ and $B$ are constants (that can be chosen arbitrarily)\footnote{Technically, these constants can be chosen such that they are imaginary numbers, but for the purposes of this discussion, we won't be making it too complex.}, and $k = \sqrt{\frac{2mE}{\hbar^2}}$. Now that it seems like we have the makings of a solution, let's think back to the original problem statement to see if we can narrow down some more things.

We're tasked with finding the wavefunction of an electron that's trapped in this infinitely deep potential well that lies between the points 0 and L. The electron will never have enough energy to pass either point---there will be no probability that we will find it outside that range. As a result, we know that the wavefunction must come to zero at at points 0 and L. We can express this as follows:

\begin{equation}
\Psi(x) = 0 \: when \: x = 0, L
\end{equation}

As a result, $B$ must be zero---if we had any nonzero constant in front of the $\cos(kx)$ term, $\Psi(x)$ wouldn't equal $0$ at the point $0$.

Now, let's think about the other bound: at L. We need the wavefunction to come to 0 at L, so:

\begin{equation}
\sin(kx) \rightarrow k = \frac{n\pi}{L} 
\end{equation}

where $n$ $\in$ $\mathbb{N}$. However, remember that we have already defined the constant $k$ when we solved our differential equation. This leads to:

\begin{subequations}
\begin{equation}
k = \frac{n\pi}{L}
\end{equation}
\begin{equation}
k = \sqrt{\frac{2mE}{\hbar^2}}
\end{equation}
\begin{equation}
\sqrt{\frac{2mE}{\hbar^2}} = \frac{n\pi}{L}
\end{equation}
\begin{equation}
E = \frac{\hbar^2}{2m}\frac{n\pi}{L}^2
\end{equation}
\end{subequations}

Once we rearrange the equation so that all of the constants are on one side, we find a strange result. When we plug in the numbers, we find that the energy that the electron can have in the well is actually limited to certain values! This is because everything except the $n$ is constant on the right side of the equation. The $n$ term, however, can hold positive integer values. As a result, the energies of the particle in the box can only have specific values. Let's try to calculate the first few for an electron trapped in a box 0.5 nanometers wide.

\begin{subequations}
\begin{equation}
E = \frac{\hbar^2}{2m} \left(\frac{n\pi}{L}\right)^2
\end{equation}
\begin{equation}
E_1 = \frac{\hbar^2}{2(9.11\e{-31})} \left(\frac{(1)(\pi)}{5\e{-10}}\right)^2
\end{equation}
\begin{equation}
E_1 = \frac{(1.055\e{-34})^2}{2(9.11\e{-31})} \left(\frac{(1)(3.14}{5\e{-10}}\right)^2
\end{equation}
\begin{equation}
E_1 = 2.41\e{-19}
\end{equation}
\end{subequations}

Plugging all of this in, we get $2.41\e{-19}$ joules. Converting in to electron volts (eV), we get that our equation is approximately equivalent to 1.5 eV for the $n=1$ energy level. Following a similar approach, we get the $n=2$ energy level to be approximately 4.5 eV. This is a very unintuitive concept. It tells us that the amount of energy in a system only comes in certain levels---you can't have any more or less than that amount. In addition, it also tells us that there is always at least some energy in a system, since the smallest value for $n$ is $1$. This lowest energy value is called the zero-point energy of the system. This idea is known as energy quantization.

Using the information above, we can finally get the general wave function for this system:

\begin{equation}
\Psi(x) = A\sin\left(\frac{n\pi}{L}x\right)
\end{equation}

\begin{figure}[ht]
\centering
\begin{minipage}[b]{0.45\linewidth}

\includegraphics[width=1.1\textwidth]{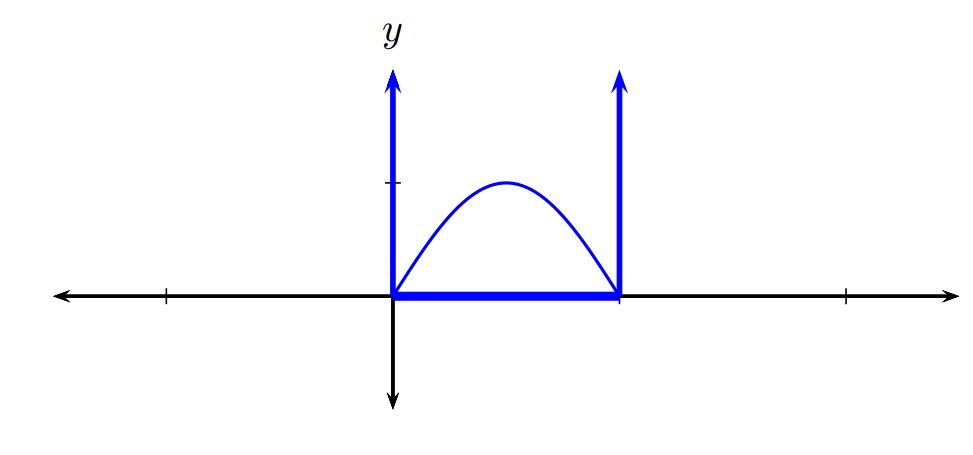}

\caption{$n$=1 next to the potential well.}
\label{fig:n1}
\end{minipage}
\quad
\hfill
\begin{minipage}[b]{0.45\linewidth}

\includegraphics[width=1.1\textwidth]{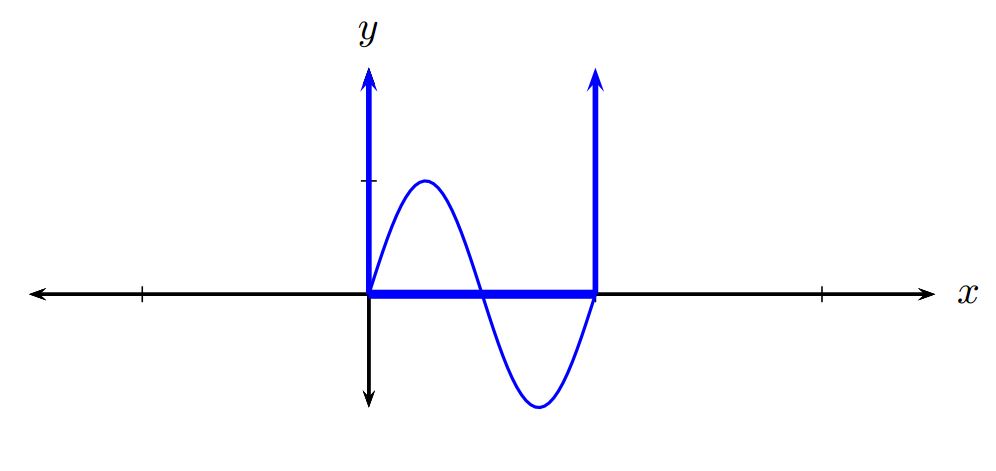}

\caption{$n$ = 2 next to the potential well.}
\label{fig:n2}

\end{minipage}
\end{figure}

You can take a look at the different wavefunctions for different energy levels in Figures~\ref{fig:n1} and~\ref{fig:n2}.\footnote{Keep in mind that the amplitude of the wave doesn't correspond the the energy of the wave---the potential energy well is just there for reference.}

As you can see, the wavefunction changes with each energy level. Because of this, the probability of finding a particle also changes with each energy level. Now, let's use the wavefunction to help us find the probability function of the particle in the box. The probability $P(x) = \Psi^2(x)$. However, we have to make sure that the all of the probabilities summed together produce 1. We can do this as follows:

\begin{subequations}
\begin{equation}
\int_{0}^{L} P(x)\,dx = 1
\end{equation}
\begin{equation}
\int_{0}^{L} \Psi(x)^2\,dx = 1
\end{equation}
\begin{equation}
\int_{0}^{L} \left(A\sin\left(\frac{n\pi}{L}x\right)\right)^2\,dx = 1
\end{equation}
\begin{equation}
\int_{0}^{L} A^2\sin^2\left(\frac{n\pi}{L}x\right)\,dx = 1
\end{equation}
\end{subequations}

A little bit of math leads to:

\begin{equation} \label{eq:normalize}
A^2 \frac{L}{2} = 1
\end{equation}

As we still haven't chosen the value of $A$, we can now set it such that the left side of Equation~\ref{eq:normalize} will equal 1.

\begin{equation} \label{eq:normalize2}
A = \sqrt{\frac{2}{L}}
\end{equation}

We now find that our probability function is:

\begin{equation} \label{eq:normalize3}
P(x) = \frac{2}{L}\sin^2\left(\frac{n\pi}{L}x\right)
\end{equation}

\begin{figure}[ht]
\centering
\begin{minipage}[b]{0.45\linewidth}

\includegraphics[width=1.1\textwidth]{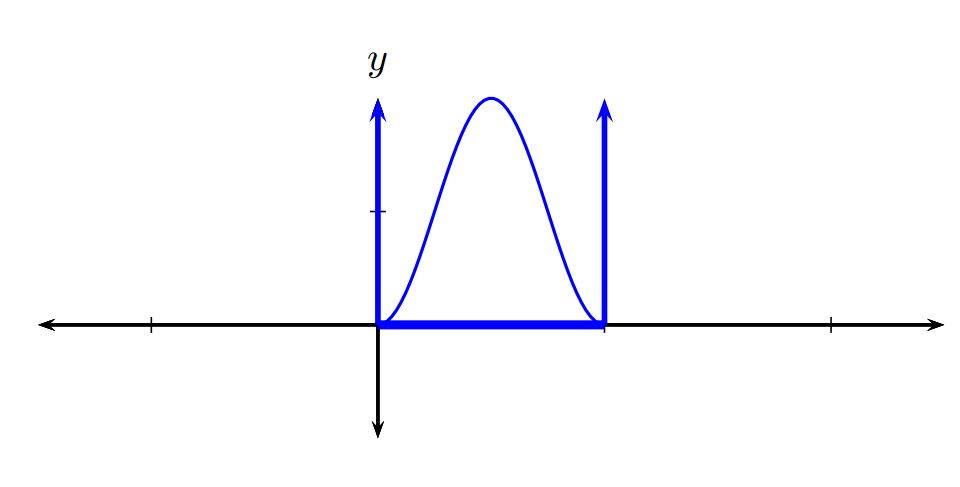}

\caption{P(x) for $n$ = 1.}
\label{fig:n1prob}
\end{minipage}
\quad
\hfill
\begin{minipage}[b]{0.45\linewidth}

\includegraphics[width=1.1\textwidth]{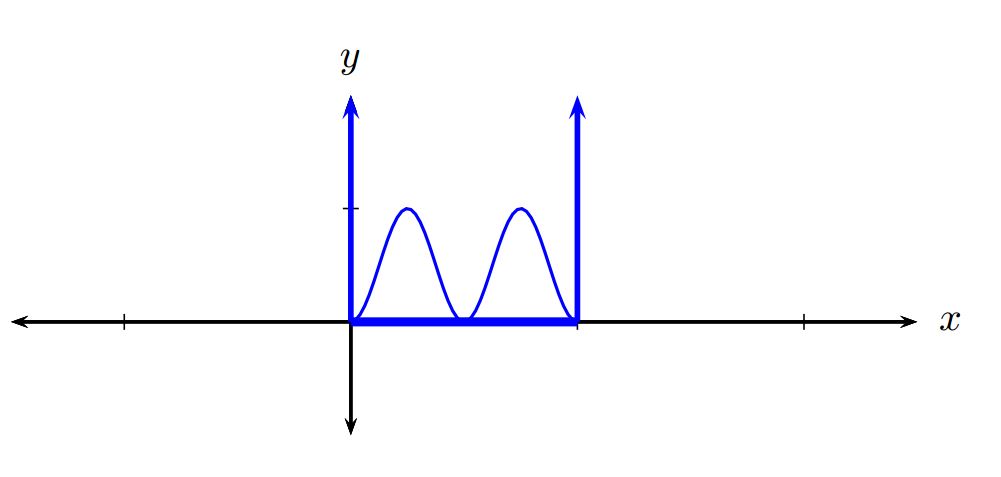}

\caption{P(x) for $n$ = 2.}
\label{fig:n2prob}

\end{minipage}
\end{figure}

This process of finding the right constants for $A$ (and $B$) is known as normalizing the wave function. You can find a visualization for the probability in Figures~\ref{fig:n1prob} and~\ref{fig:n2prob}.

Once again, we see some strange results. For the $n = 1$ energy level, we find that the particle is more likely to be found in the middle of the well rather than the sides. Looking at the $n=2$ energy level, we find that it acts strange in other ways. Namely, we can never find the particle in the middle of the box---the probability there is zero. 

Once again, we find that this makes no sense whatsoever intuitively. Why would the predicted position change if we add just a little bit of energy to the particle? No one knows. After all, this is quantum physics: it's impossible to get an intuition for.

\section{Conclusion}

It's important to understand the position that math carries in the field of quantum physics. Math gives us a window to glimpse the truth in this field, but after all, it's only a window. Suppose the Wave-Particle Duality, for example. In an ordinary laboratory experiment, it's possible to measure a particle, but not a particle's wave. How do we know it's there, then? We carry out another experiment which shows a different behavior than than which would have been expected if the object had acted like a particle.

At this point, we can take the evidence of the discrepant behavior and see if it matches up with anything else. And it does. By modeling the particle as a mathematical wave, we can get the behavior that suits the results of the experiment.

We now have the results that we want, but we've gotten them in a very dangerous way. We've ``modeled'' the particle as a wave, but we have \textit{no} experimental evidence that this is true. At this point, we're using math as a crutch to support what we think is true about the world. 

This isn't a very safe place to be in for the subject of physics, which is, at heart, an experimental science. However, as scientists get deeper and deeper into the field of quantum physics, we will start to understand the true beauty of the universe as it was, is, and always will be.

\end{document}